\documentclass[]{aa}

\usepackage[varg]{txfonts}
\begin{document}


\newcommand{\be}{\begin{equation}}
\newcommand{\ee}{\end{equation}}
\newcommand{\bea}{\begin{eqnarray}}
\newcommand{\eea}{\end{eqnarray}}
\newcommand{\nn}{\nonumber}
\newcommand{\pgb}{\pagebreak}
\newcommand{\noi}{\noindent}
\newcommand{\appgeq}{\stackrel{>}{\sim}}
\newcommand{\appleq}{\stackrel{<}{\sim}}
\newcommand{\etal}{ {\it et al.} }
\newcommand{\mt}{\bf}


\title{\bf  Galactic Magnetic `X  Fields'}
\author{R.N. Henriksen }
\institute{Dept. of Physics, Engineering Physics \& Astronomy, Queen's University, Kingston, Ontario, K7L 3N6, Canada}
\date{ Received YYY/Accepted XXX}
\abstract

{Our aim in this note is to compare a recent explanation of the galactic X pattern in Faraday Rotation (XRM), to that produced by the advection part of the classical dynamo.}{We find that the characteristic X magnetic field polarization in the plane of the sky, found in edge-on spiral galaxies, can develop magnetohydromagnetically  from an initial disc magnetic field combined with wind and rotation.  The Rotation Measure  develops a corresponding X distribution in sign, but this distribution is not a `universal' behaviour because it depends primarily on the velocity field. We use Cauchy evolution of an initial magnetic field to find the field at some later time.} {A `battery' mechanism that requires current to always flow out of a galaxy has been recently suggested, which contrasts with the conclusions of this paper. }  {Either explanation has significant consequences for the structure of a spiral galaxy. If the battery mechanism applies then we have a new method of producing magnetic field independent of the traditional dynamo. If however the flow mechanism applies; then constraints concerning  the presence and nature of a galactic wind, together with the signature of the mean radial magnetic field,  can be inferred. }{}

\maketitle



\section{Introduction}

Recent work by the CHANG-ES collaboration \cite{KIS2020} has demonstrated, by stacking the maps of the sample galaxies, a certain universality of the `X' magnetic field topology in edge-on spiral galaxies. This X topology refers to the polarization structure in the plane of the sky, which forms an `X' centred on the galactic centre.This result has been expanded in an interesting paper \cite{MC2021} that has  presented evidence for an X type pattern  in the  sign of the Faraday Rotation Measure (XRM)\footnote{Here the X pattern refers to the oscillating sign of the longitudinal field in each quadrant. See sketch.}. The sense of the implied toroidal magnetic field is such as to give outward current along the axis of a galaxy. The authors interpret this in terms of a `battery mechanism', introduced in \cite{CK1998}, rather than as the result of an magnetohydrodynamic (MHD) `dynamo'. The mechanism produces current parallel to the galaxy rotation velocity, that is, always outwards from the galaxy. 

 The observational evidence \cite {MC2021}, although indicative, is not yet definitive \cite{Beck2021}. For example it would be better not to mix rotation measures (RM) in L band with those measured in C band when performing the stacking. The signal to noise ratio of the field polarization for each of the galaxies should be taken into account. The inclusion of galaxies possessing a strong central radio source favours the new mechanism, but may also reduce sensitivity to the weaker structure. It would be better to have a larger sample because the evidence for this new result depends strongly on only  few galaxies. For these reasons we present  in this note an alternative explanation for the XRM pattern, based on advection of a magnetic field produced by a `fast dynamo'.     

Because the battery mechanism depends on radiation from a central source interacting with electrons in an accretion disc, we may expect it to be most effective in the central regions of the galaxy. This is roughly in accord with the findings reported in \cite{MC2021}, although the effect seems to persist out to nearly half the radius of the galactic disc. In this same region  of an edge-on galaxy one finds frequently X magnetic field structure and X structure in the Faraday rotation measure(XRM), due wholly to dynamo theory \cite{HI2016},\cite{Hen2017}, \cite{WHIM-P2019}.  In fact those papers may be said to predict  a XRM behaviour, although it is not universal in the sense of \cite{MC2021}. 

There are various combined effects included in the general  dynamo theory including advection, rotation, diffusion and a sub-scale fast dynamo ($\alpha$ effect, \cite{SKR1966}). It is our intention in this paper to reduce these effects to the bare minimum necessary for the X magnetic field and the XRM configurations.  We show that with these minimal elements  a similar region of the galaxy can show both configurations.

In detail we show that an advection/rotation flow, working magnetohydrodynamically  (MHD) on an '$\alpha$ dynamo' disc magnetic field \cite{BZK2017} that we assume, can produce the same X topology both in the polarization and the RM. The X field structure and the toroidal field were already apparent in \cite{HI2016}, using a self-similar model of galactic outflow. However, unlike the case with the battery mechanism, the sense of the toroidal field and hence of the axial current can vary, because it depends on two incidental factors. 

One factor is  the sign of the radial component of the initial disc magnetic field. In particular, we find that it is necessary that the sign of this component change on crossing the disc in order to establish the XRM topology.   Such field components may arise in axially symmetric dynamo theory \cite {KF2015}, \cite{HI2021}.The second factor is the sense and radial variation of the galactic rotation. These two factors interact.

 In our example we have used  left handed galactic rotation and a positive radial magnetic field `above' (i.e. that is where the angular velocity vector points  out of the disc). This gives the same RM distribution suggested in \cite{MC2021}. However the same topology arises with right handed rotation and a negative radial field above the disc. If only one of the rotation or the radial field is reversed, then the sense of the toroidal field is such as to have inward flowing current.
 
 In addition to the wind velocity increasing with radius, the X field behaviour  depends on the angular velocity decreasing with radius as in equation (\ref{eq:rot1}) of the appendix. Therefore we exclude a region of rigid body rotation  in favour of differential rotation, which restricts our model to intermediate scales. One can change the sense of the toroidal field in the modelled region by changing the sign of either the radial field or the angular velocity.
 
 Should we have right handed rotation together with  a disc radial field positive above the disc, then we predict the RM to be wholly positive on one side of the axis and wholly negative on the other, so that there is no XRM configuration although there is likely, X magnetic field. The left/right signs are interchanged given left handed rotation and an inward radial field in the disc.


Hence, our model requires that there should {\it not} be universality of the sense of the XRM  topology and hence of the  current direction on intermediate (i.e. kpc) galactic scales. Smaller scales (perhaps at less than $1$ kpc) may be more subject to the battery mechanism, although we would expect an AGN jet to produce similar topology \cite{HI2021}. 

In this note we make use of the Cauchy integral for the magnetic field as a function of time and place; given an initial magnetic field and a prescribed wind and rotation, each  of these with a spatial gradient . Such an advected magnetic field  is only one element of a classical magnetic dynamo (e.g. , \cite{SKR1966},\cite{M1978}, \cite{BS2005}, \cite{KF2015}) and it can not by itself be regarded as a self-sustaining dynamo, because of the assumed lack of back reaction on the flow due to induced currents. However given an initial magnetic field and a velocity field of high magnetic Mach number, the Cauchy solution describes the exact transitory development of the magnetic field until energy equipartition with the flow is achieved.

 An initial magnetic field implies a pre-existing field source, probably a sub-scale  turbulent (i.e. an $\alpha$ dynamo) dynamo (e.g. \cite{BZK2017} for supernovae driven turbulent dynamo).  We assume an initial field that is parallel to the galactic disc and radial, for clarity in following the advected development. This does not exclude the presence of other components of the field in a realistic initial condition. It should be noted that the turbulence in the $\alpha$ dynamo is below the scale on which the global field appears. The global behaviour is due to the assumed smooth variation of the $\alpha$   
parameter function.

This description would cease to apply in a closed system as the magnetic field energy came into equipartition with the flow energy. In an open system  such as may be a galactic halo, the magnetic flux can be steadily lost to the intergalactic medium. It might be  replenished  by the alpha dynamo in the disc faster than it is advected away. This would depend on the supernova rate and the wind transit time.

If the flow has internal spatial gradients, these act to create  a new magnetic field in a kind of advection dynamo.  A simple example of amplification or dilution behaviour in time and space for each field component is given briefly in appendix (\ref{sect:Theory}). This accounts for the increasing field strength in our models with height and time. Once again this must stop, once the field energy density come to dominate the flow energy density. 

We remove the technical details, including definitions and physical laws, to the appendix. There, the Cauchy integral for an accelerated vertical  wind with rotation is used in cylindrical coordinates based on the galactic axis. This integral follows from the continuity equation plus the usual MHD equation
 that assumes zero resistivity. Because it is a Lagrangian method, it allows the advected magnetic field to be found once the displacement field is given. No integration of the differential equations is required. The advection occurs in the classical dynamo theory in combination with the $\alpha$ effect  and diffusion. This flow gives the $\alpha/\Omega$ effect when differential rotation exists. We ignore diffusion in favour of galactic wind, and assume that the $\alpha$ effect has produced our initial field.
 
\section{Examples}

 Scales in space and time are used in the Appendix (\ref{sect:Theory}), and appear as $\{V_N,\Omega_d, \varpi_N, z_d\}$  that may be  freely chosen.  The height $z_d$  is a vertical scale in terms of which all other heights are measured.  The velocity $V_N$ is the wind velocity at cylindrical radius scale $\varpi_N$, and $\Omega_d$ is the angular velocity of a particle initially at $z=z_d$.  The height $z_d$ is roughly the height at which the wind begins. Together these define  the time from the initial configuration, the wind and angular velocities, and the spatial scales in height and radius. The subscript `N' suggests nuclear values, while $z_d$ is essentially half the thickness of the galactic disc.

Additional constants are defined in the Appendix but for convenience they are:
\bea
\omega&=&\Omega_d\frac{z_d}{V_N},~~t\leftarrow t\frac{V_N}{z_d}, ~~\varpi_o\leftarrow \frac{\varpi_o}{\varpi_N},\nn\\
z_o&\leftarrow& \frac{z_o}{z_d},~~\beta =\frac{B_{oz}}{|B_{o\varpi}|}, ~~\beta p=\frac{\varpi_N}{z_d}\beta,. \label{eq:units1}
\eea
 In addition the field strength is given in terms of the initial value of the radial magnetic field  $B_\varpi(0)$.
The subscript $o$ indicates a value at an initial position in the rotating outflow, namely at $\{\varpi_o,z_o\}$. We take $B_{oz}=0$ for simplicity in this example, although this is not necessary for the conclusion. There is an additional parameter $k$ that defines how fast the radial velocity at an initial point falls off with radius, given in equation (\ref{eq:MapZradial}).  In addition to the initial field strength, numerical values in our calculations require the numerical assignment of the constant $V_N$ for the velocity, $\varpi_N$ for the radius, $z_d$ for the halo scale and $\Omega_d$ for the angular velocity. 
These can be chosen appropriately for any galaxy of interest.

\begin{figure*}{}
\begin{tabular}{cc} 
\rotatebox{0}{\scalebox{0.4} 
{\includegraphics{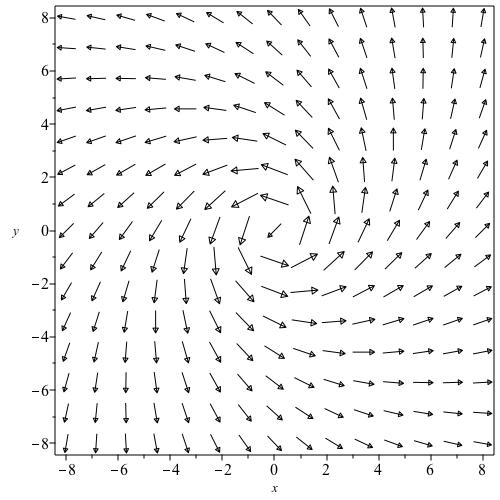}}}&
\rotatebox{0}{\scalebox{0.4} 
{\includegraphics{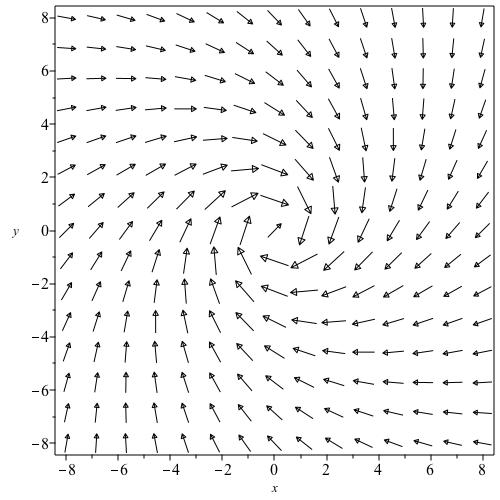}}}\\
{\rotatebox{0}{\scalebox{0.4} 
{\includegraphics{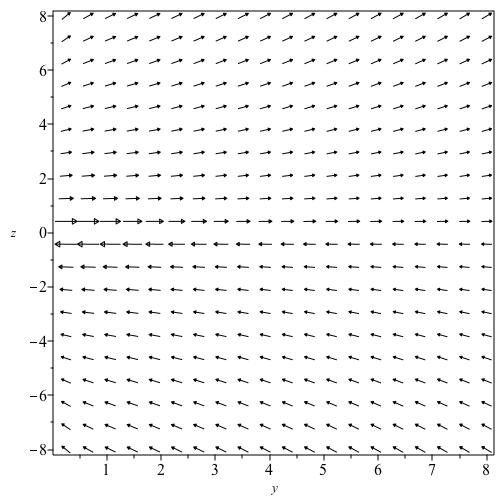}}}}&
\rotatebox{0}{\scalebox{0.4} 
{\includegraphics{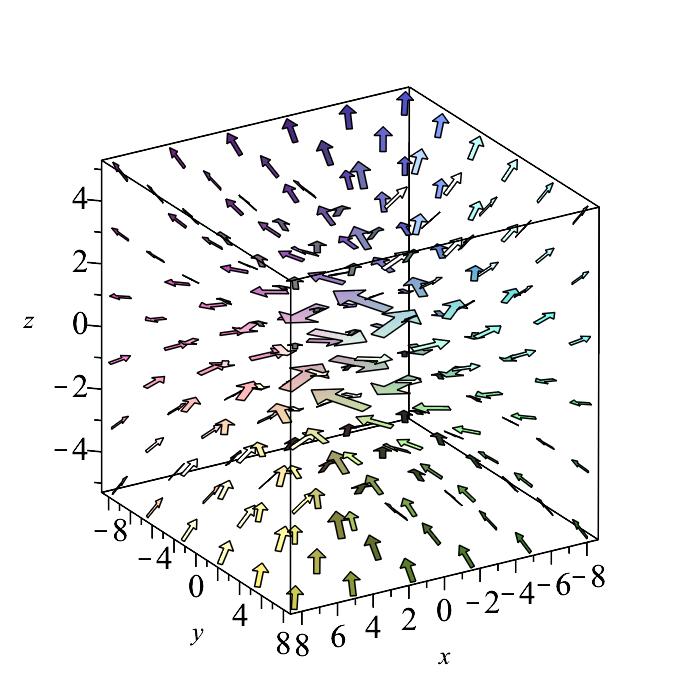}}}
\end{tabular}
\caption{ This figure plots the Cartesian components of the magnetic field that is axially symmetric in cylindrical coordinates.The upper left panel shows the magnetic field in the $\{x,y\}$ plane cut at $z=+1$. The upper right panel shows the $\{x,y\}$ plane at $z=-1$.
the lower left panel shows the $\{y,z\}$ poloidal plane at $x=1$.  The lower right panel is a 3D view of the field above and below the plane. The parameters used in the figure are $\{k,s,\beta,\beta p,\omega,t\}=\{-0.03,1,0,0,-2,4\}$.  The scales are $\varpi_N$ for the radius and $z_d$ along the axis. The arrows show the direction of the field at the end point of the arrow, but the field strength in the first three figures has been adjusted so as to show the outer arrows clearly. The fourth figure shows each arrow length as a fraction of the maximum value. }

\label{fig:3Dfield}
\end{figure*}

\begin{figure*}{}
\begin{tabular}{cc} 
\rotatebox{0}{\scalebox{0.3} 
{\includegraphics{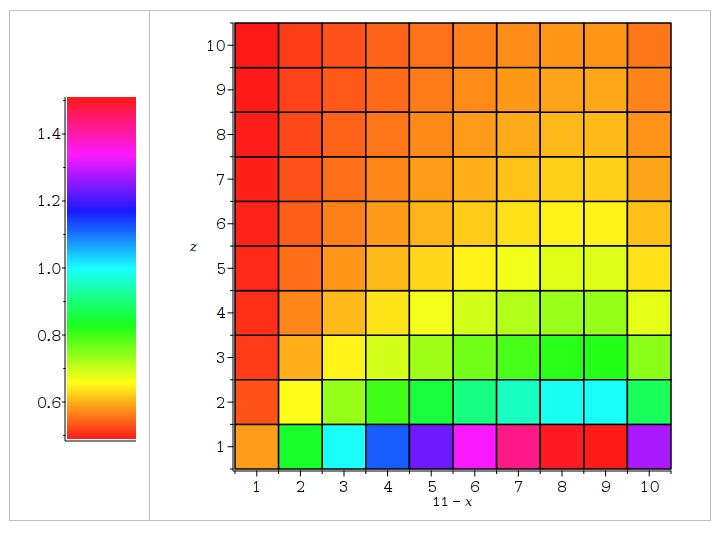}}}&
\rotatebox{0}{\scalebox{0.3} 
{\includegraphics{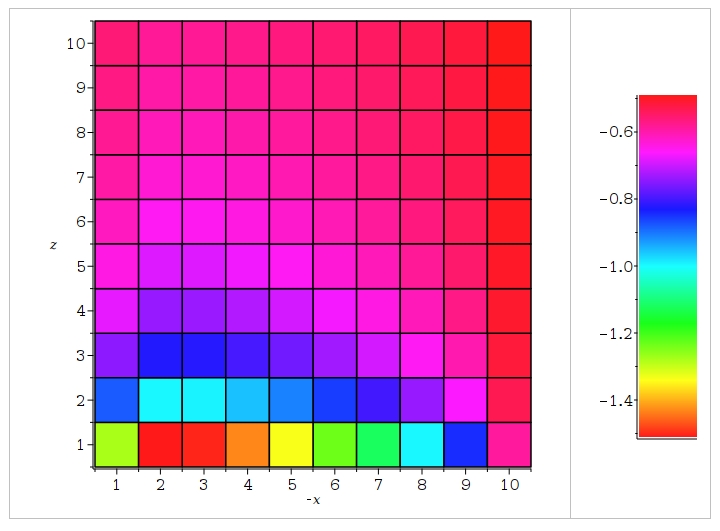}}}\\
{\rotatebox{0}{\scalebox{0.3} 
{\includegraphics{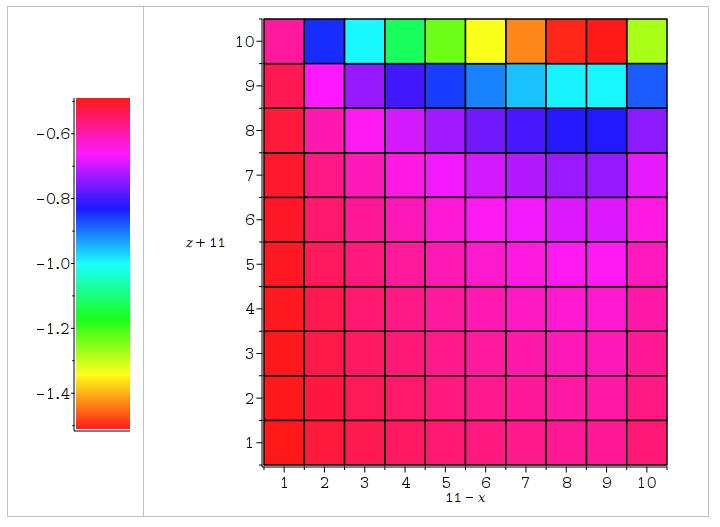}}}}&
{\rotatebox{0}{\scalebox{0.3} 
{\includegraphics{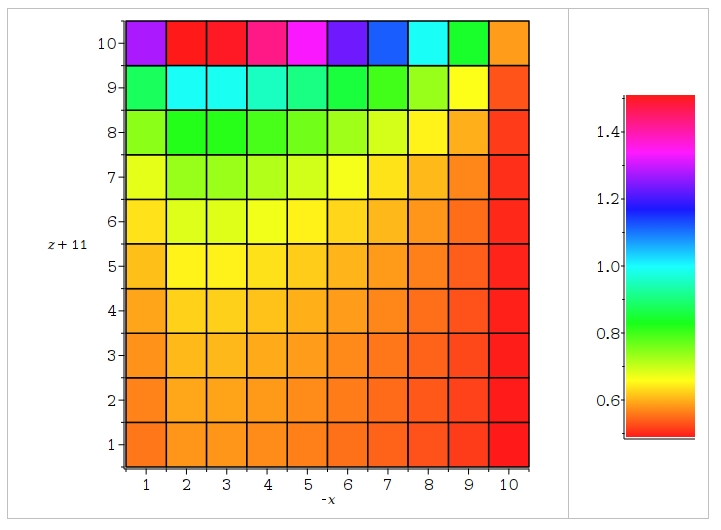}}}}
\end{tabular}
\caption{This figure shows the distribution of the integrated line of sight RM in the four quadrants on the sky determined by the galactic edge-on disc and the  positive rotation axis. We are looking along the  negative y axis relative to figure (\ref{fig:3Dfield}).  The parameters of the magnetic field and of the flow field including rotation are as in  the upper row of figure (\ref{fig:3Dfield}).  The upper right and lower left images contain only negative RM values while those at upper left and lower right are positive.  We see that the sign changes across the disc  and across the axis to give a characteristic `X configuration' in the RM. The positive behaviour is at upper left and lower right while the negative behaviour is at upper right and lower left.}
\label{fig:compositeRM}
\end{figure*}

\begin{figure*}{}
\begin{tabular}{cc} 
\rotatebox{0}{\scalebox{0.6} 
{\includegraphics{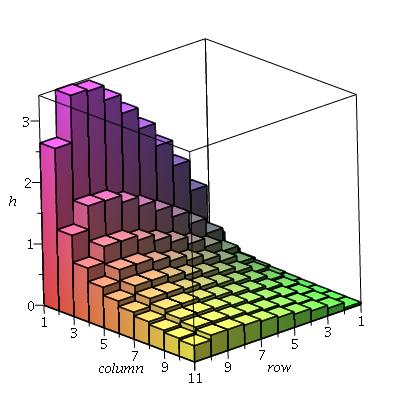}}}\\
\end{tabular}
\caption{ This diagram gives the value on each patch of the Rotation Measure in the second quadrant of our example. Other quadrants are the same except for sign.  The colour is merely shading and has no physical significance.To ensure that we are looking along the negative y axis  across the positive x axis, 
the row is $11-x$ and the column is the $z$ axis. The observer is looking down from the left.}
\label{fig:valuesRM}
\end{figure*}

Figure (\ref{fig:3Dfield}) shows an evolved magnetic field above and below the galactic disc, provided the radial field changes sign across the disc. The figure is axially symmetric in cylindrical coordinates but Cartesian components are shown. 
The principal message from  the upper panels of this figure is the spiral nature of the field in this `cross cut', whose winding sense  reverses across the equatorial plane. The panels show the $\{x,y\}$ planes at $z=\pm1$. The observer is looking perpendicular to the $z$ axis so that when a number of these planes are stacked in $z$, an extended line of sight magnetic field appears that reverses direction above and below the plane and across the axis. This reversal of field direction below the plane and across the axis corresponds to the  reversing RM pattern shown in figures (\ref{fig:compositeRM}) and (\ref{fig:valuesRM}). This  field evolution uses the flow field described in (\ref{sect:kinematics}). 

The lower left panel shows the $\{y,z\}$ plane in a cut at $x=1$. The strength of the field falls off faster form the axis than is indicated by the length of the arrows
 Such a field projected onto the sky will show an X field about the axis, due to polarization generated mainly where the field is most transverse to the line of sight (los). The field closer to being along the los will generate the characteristic RM pattern or its negative. 

Finally the panel at lower right shows the local directions of the field in a full 3D space. The coordinates are also Cartesian but the field strength is a fraction of the maximum as indicated by the length of the arrows. Near the plane, the sense of field rotation and the sign of the radial field component
are seen to reverse.

Figure \ref{fig:compositeRM} illustrates  the  XRM  pattern produced by the  toroidal component of the field for the same parameters as in the upper row of figure (\ref{fig:3Dfield}). We are still looking along the negative y axis as defined in figure \ref{fig:3Dfield}, so {\it  x} increases positively to the left and negatively to the right.  This X pattern in the RM signature  evolves from the initially radial field {\it that changes sign across the galactic disc}. If the initial radial field does not change sign on crossing the disc, the pattern would be wholly negative on one side of the axis and wholly positive on the side across the axis. 

The signs of the RM in the pattern can be interchanged either by reversing the sign of the radial field component above and below the disc, or by reversing the sense of the galactic rotation. We have used left handed rotation about the $z$ axis in these examples.

It should be emphasized that this XRM pattern is the inevitable result of an $\alpha$ small-scale turbulent `fast' dynamo plus a galactic wind and halo rotation.  The same behaviour may be demonstrated if the wind moves in spherical radius with a $\cos{(\theta_o)}$ co-latitude dependence. A supernovae driven nuclear wind was already studied in (\cite{RC1985}). 

The wind and rotation  together cause  the small-scale magnetic field to be expanded in scale. The advection is a kind of macroscopic diffusion.  The importance of the advection confirms a recent conclusion (\cite{HI2021}) based on scale invariant dynamo theory.

 Our `Rotation Measure' is calculated as the line integral of the line of sight magnetic field in each quadrant, using the convention of positive field towards the observer. As such only the signs plus relative values are significant. It should be noted that a similar pattern is quoted for the Milky way in \cite{OJR2012}. Indeed the inferred quadrupole does possess the pattern required in \cite{MC2021}, but it is on a very large scale.

In figure (\ref{fig:valuesRM})  a discrete block diagram of the RM values in the second quadrant is shown. The values are the same in the other quadrants but for a negative sign in the first and third quadrants. The height of the diagram in each block gives the  mean RM value in the block. The blocks correspond to the grid divisions in figure (\ref{fig:compositeRM}), where the values are smoothed. The label `row' stands for $11-x$ and `column' stands for $z$.

\section{Discussion and Conclusions}

 In this paper we have identified a flow field and an initial magnetic field that  together produce the characteristic X magnetic field, which has proven common in edge-on spiral galaxies (see e.g. \cite{KIS2020} and references therein). The basic requirement is an initial magnetic field that has a radial component, which changes  sign across the disc \footnote{That is a  quadrupole topology, because the vertical field vanishes. This allows for a change in sign of the radial field across the disc because there is no field connection through the disc.}, plus a galactic wind that increases outwards. 
 
 To achieve a characteristic `X distribution' of the rotation measure (RM) the wind must have rotation.  We have allowed this rotation  to have a `lag' and the vertical velocity to accelerate. Both of these effects enhance the strength of the magnetic field but are not essential for the two X behaviours.  
 
 It is interesting  to note that these elements are present in recent series of studies on the asymptotic galactic dynamo (see e.g. \cite{HI2021}) and references therein), which also produce the `X field' topology.  The fast sub-scale dynamo ($\alpha$ effect) is also present in those studies as is also in some cases diffusion.  However the $\alpha$ effect is primarily only necessary for the self-consistent production of the field on which the flow works. The field `relaxation' produced by diffusion is dominated by a galactic outflow when present.
 
 Our important conclusion relative to the literature is that, unlike the recent interesting suggestion by \cite{MC2021}, the sense of the X pattern in the Faraday Rotation is not necessarily such that current flows outward along the axis of the galaxy. By changing the sense of galactic rotation, or by reversing the sign of the radial magnetic field , our mechanism produces a pattern that corresponds {\it either} to outward or inward axial current flow. Should the sign of the radial field be the same on both sides of the disc, then the X RM pattern disappears in favour of a simple sign change across the disc.

\begin{figure}{}
\rotatebox{0}{\scalebox{0.4} 
{\includegraphics{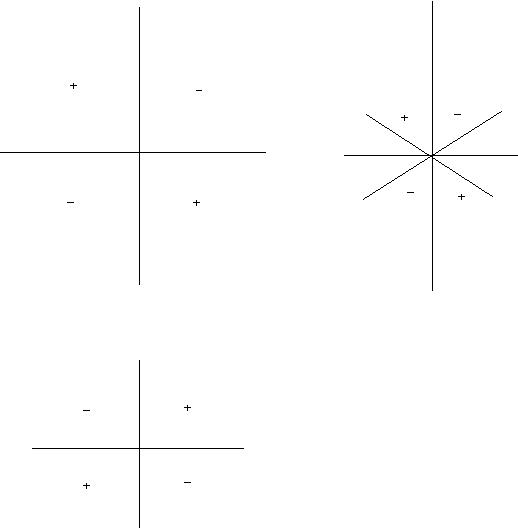}}}
\caption{At left we sketch the two possibilities for the RM signature in each quadrant. A plus sign indicates a field approaching the observer.  Hence the upper image indicates outward current along the rotation axis and the lower image indicates inward current. The sketch at right shows the projected polarization that comprises an `X' field. The signs indicate the position angle relative to the rotation axis, if the angle is measured positively counter clockwise. The actual angles vary considerably both in theory and observation. }
\label{fig:sketch}
\end{figure}

The two possibilities admitted by this paper are shown, in terms of the sign of the RM in each quadrant, in the left column of the sketch (\ref{fig:sketch}). The upper left  implies current flowing outwards along the axis while the lower left has inward flowing current. We admit both possibilities in this paper, while only the upper left is possible with the battery mechanism. At right the projected `X' polarization is sketched. The signs indicate the position angle relative to the rotation axis, measured positively counter clockwise.

This means that we have a clear contrast between our radially increasing  wind with rotation acting on a radial field component and the `battery mechanism' of \cite{MC2021}. The stacked RM for magnetized spiral galaxies should NOT, according to our advection mechanism, show a strong tendency to the configuration given by outward axial current. In \cite{MC2021}, evidence was presented that such a stacked image does yield the electromagnetic current outflow configuration required by the battery mechanism.  We must however await a definitive conclusion when improvements mentioned previously are included. This model can also be tested in individual galaxies (even for the Milky way, \cite{OJR2012}) by measuring both the angular velocity and the RM in four quadrants, but we do not predict stacking of the RM to show any strong residual signal. One can only anticipate more studies of this type, before a conclusive answer will be available. It is rare to have such a conclusive test of two contrasting possibilities.

\section{Acknowledgements}

This paper has had the advice of my wife Judith Irwin, for which I am grateful. The referee, Rainer Beck, provided much constructive criticism.

\begin{appendix}
\section{ Basic Theory}
\label{sect:Theory}
\subsection{Basic assumptions}

We recall that formally the Cauchy solution for the magnetic field given a velocity field is 
\be
{\bf B}=\frac{\rho}{\rho_o}({\bf B}_o\cdot \nabla_o){\bf R}.\label{eq:Cauchy1}
\ee
Here a subscript $`o`$ indicates values at an initial point in the flow. The gas density is $\rho$, ${\bf B}$ is the vector magnetic field and ${\bf R({\bf r},t)}$ is the current location of the initial point. The current density is given in terms of the initial density by

\be
\frac{\rho}{\rho_o}=\frac{1}{J({\bf R},{\bf r})},\label{eq:density1}
\ee
where $J$ is the Jacobian between the new (${\bf R}$) and old (${\bf r})$ coordinate locations. We calculate ${\bf R}$ in appropriate coordinates generally  as 
\be
{\bf R}={\bf r}+\int~{\bf v}({\bf R},t)dt,\label{eq:LagrangeR}
\ee
but we will choose a simpler form for the velocity field than this expression assumes.

In Cartesian coordinates the meaning of these various expressions is clear, but in curvilinear coordinates a little care is required. We let the {\it physical} increments along the current and initial curvilinear axes to be designated $dR_i$ and $dr_i$ respectively. In cylindrical coordinates $\{\varpi,\Phi,Z\}\equiv \{1,2,3\}$ with the $3$ axis coincident with the rotation axis of the galaxy, these become 
\bea
dR_1&=&d\varpi,~~~~dR_2=\varpi d\Phi,~~~~dR_3=dZ,\nonumber\\
dr_1&=&d\varpi_o,~~~~dr_2=\varpi_o d\phi,~~~~dr_3=dz.\label{eq:physincrements}
\eea
This allows us to write the solution (\ref{eq:Cauchy1}) in a practical form for curvilinear coordinates 
\be
B_i=\frac{1}{J}\frac{\partial R_i}{\partial r_j}B_{oj},\label{eq:Cauchy2}
\ee
where the summation convention is used. The Jacobian becomes in these terms 
\be
J=det\big(\frac{\partial R_i}{\partial r_j}\big).\label{eq:jacobian1}
\ee

 Equation (\ref{eq:Cauchy2}) gives each field component in terms of a `stretching matrix' ($\partial R_i/\partial r_j$) operating on the initial magnetic field multiplied by the change in density. This operation amplifies the field in an accelerating wind, or under differential rotation, provided that  the density does not fall off too rapidly (cf equation \ref{eq:jacobian1}). It is well known, and can be found recently as equation (3.28) in \cite{BS2005}.
 
  The simplest example of combining our flow field with equation (\ref{eq:Cauchy2}) gives a mixture of amplification and dilution. Thus, a field at a current position that began at $\{\varpi_o,z_o\}$ will have after a time $t$ the components (the radial velocity $u$ is given by equation (\ref{eq:MapZradial}) below)
\bea
B_\varpi&\propto &\frac{1}{1+u(\varpi_o,k)t},\nn\\
B_\phi&\propto& -\frac{1}{1+u(\varpi_o,k)t}\big(\frac{\varpi_o^2}{z_o(1+\varpi_o^2)^{3/2}}\omega t\big),\nn\\
B_Z&\propto& -\frac{kz_ou(\varpi_o,k)t}{1+u(\varpi_o,k)t}.
\eea
The proportional sign allows for an arbitrary strength of the initial radial magnetic field so the quantities on the right are local amplification/dilution factors. We see that due to the density dilution the radial field decreases in time, and with increasing radius. The azimuthal field stabilizes in time at $\omega/(u(\varpi_o,k)z_o)$ but decreases  with radius and height. The vertical field stabilizes in time and radius, and increases with height. The flow field of the following section can be used to convert from initial position to current position.

Once we insert into these formulae a suitable form from equation (\ref{eq:LagrangeR}), the magnetic field is known as a function of time and the initial coordinates. For this we must choose the form of the velocity flow field that establishes the coordinate transformation ${\bf R({\bf r},t)}$. This is the subject of the next section.

\subsection{Flow field}

\label{sect:kinematics}

We will treat the simplest possible wind in cylindrical coordinates. Throughout we assume axial symmetry. The wind outflow takes the form 
\be
{\bf v}=\hat {\bf e}_z u(\varpi_o)z_o,\label{eq:windvel}
\ee
so that it is purely vertical and a function of its initial radius and height. We consider it to be established as a constant quantity by the galaxy. 
  We choose  the wind velocity to be proportional to the initial height $z_o$ rather than to  the current height $Z$. Therefore each Lagrangian wind element  labelled $z_o$ moves with a constant velocity, whose value is fixed by the velocity at the initial height of the element. This initial velocity increases linearly with $z_o$ (taking the time derivative of equation \ref{eq:MapZ}). 

Consequently, we have the transformation from initial vertical coordinates of a flow element to current vertical coordinates of the flow  element as
\be
Z=V_N u_o(\varpi_o)z_ot+z_o,\label{eq:MapZ}
\ee
where $V_N$ is the amplitude of $u(\varpi_o)$.  The scale velocity  $u_o(\varpi_o)$ is dimensionless. We will take this  radial function to be
\be
u_o(\varpi_o)=e^{k(1-\varpi_o)},\label{eq:MapZradial}
\ee
where $\varpi_o$ is now measured in units of  a radial scale $\varpi_N$, at which radius the vertical velocity is equal to $V_N$.  The numerical constant $k$ allows the scale of the exponential decline in outflow to be adjusted. If this value is close to $1$, and if $\varpi_o$ is small, this form  may imitate the rapid decline from a nuclear jet. However a key result of this paper is that $k$ should be negative and small in order to produce a definite X field. This describes a wind velocity that increases with cylindrical radius in the galaxy.

The rotation velocity  is taken to have the form
\be
v_\phi=\frac{\Omega(\varpi_o,z_o)\varpi_o}{\sqrt(1+\varpi_o^2)}=\varpi_o \dot\phi,\label{eq:rot1}
\ee
where $\dot x\equiv dx/dt$. The radius $\varpi_o$ does not change during the flow, but we will allow for a moderate `halo lag' (\cite{R2000}, \cite{Heald2009}) 
as the flow element moves vertically by taking
\be
\Omega=\frac{\Omega_d z_d}{|z_o|}.\label{eq:omega}
\ee
Here the constant $\Omega_d$ is equal to $\Omega$ at $z_o=z_d$ and $t=0$. We choose the halo lag to depend on the initial vertical location of the  wind element because we take this to be set by the larger galaxy. This gives a rotation angle that varies linearly in time.
The form  (\ref{eq:rot1}) gives a flat galactic rotation velocity at large $\varpi_o$ and a linear rise at small $\varpi_o$. One notes that the angular velocity  $\dot\phi$ increases to smaller $\varpi_o$.

Equations (\ref{eq:MapZ}), (\ref{eq:rot1}) and (\ref{eq:omega}) can now be combined to give  

\be
\phi(\varpi_o,z_o)=\phi_o+\frac{\omega}{z_o}\frac{t}{\sqrt(1+\varpi_o^2)}.\label{eq:azimuthincrease}
\ee

We have defined the dimensionless quantities
\be
\omega=\Omega_d\frac{z_d}{V_N},~~~~t\leftarrow t\frac{V_N}{z_d}, ~~~~\varpi_o\leftarrow \frac{\varpi_o}{\varpi_N},~~~~z_o\leftarrow \frac{z_o}{z_d}. \label{eq:units}
\ee

Using the definition (\ref{eq:jacobian1}) and the mappings (\ref{eq:MapZ}) (scaled by $z_d$), (\ref{eq:azimuthincrease}), and $\varpi=\varpi_o$ we find that 
\be
J=\frac{\partial Z}{\partial z_o}=1+u_o(\varpi_o)t\equiv \frac{\rho_o}{\rho}.  \label{eq:Jacobian}
\ee

Equation (\ref{eq:Cauchy2}) gives finally the magnetic field as a function of time in the explicit form for cylindrical coordinates $\{\varpi,\phi,z\}$ as 
\bea
B_\varpi&=&\frac{B_{o\varpi}}{J},\nonumber \\
B_\phi&=& \frac{1}{J}(\varpi_o\frac{\partial \phi}{\partial \varpi_o}B_{o\varpi}+\varpi_o\frac{\partial\phi}{\partial z_o}B_{oz}),\label{eq:magneticfield1}\\
B_Z&=&\frac{1}{J}(\frac{\partial Z}{\partial \varpi_o}B_{o\varpi}+\frac{\partial Z}{\partial z_o}B_{oz}).\nonumber
\eea
The necessary derivatives follow from equations (\ref{eq:azimuthincrease}), (\ref{eq:MapZ}) plus the choice of $u(\varpi_o)$ in equation (\ref{eq:MapZradial}).

We can use these equations to find the magnetic field development as a function of time over  the meridian plane. This development is studied in the text. 

We scale the magnetic field strength by $|B_{o\varpi}|$ in our sample calculations. Because this equals  $A/\varpi_o$ for some constant $A$, the calculated fields must be multiplied by $A/\varpi_o$. This assumes that the initial field is replenished by the galaxy on a time scale that is short compared to the advection time. This can be altered by taking the initial field to depend on $\varpi(\varpi_o,t)$ so that the outflow is no longer cylindrical, but that is another model.

Because of the modulus, if the initial radial field (i.e. a `split monopole') reverses direction across the galactic plane  $B_{o\varpi}$ should be replaced by $signum(z_o)$ in equations (\ref{eq:magneticfield1}). The length and time scalings are as discussed above. The derivative $\partial \phi/\partial z_o$ generally requires a $signum(z_o)$ factor at $z_o<0$ so that $\phi$ may decline with increasingly negative $z_o$. If we take the initial magnetic field to lie wholly in the meridional plane, then $B_{o\varpi}\propto 1/\varpi_o$ and $B_{oz}$ must be constant in $z$, in order that the field divergence vanishes. In fact $B_{oz}=0$ in our simple model.

Formally we introduce  constants $\beta p$  and $\beta$ given by 
\be
\beta =\frac{B_{oz}}{|B_{o\varpi}|}, ~~~~~\beta p=\frac{\varpi_N}{z_d}\beta,\label{eq:constants}
\ee
but in this study they are both set equal to zero.

We have reduced the model to a toy model by our assumptions, in order that one principal effect  (edge-on X magnetic field) of a radially varying  galactic disc wind  and galactic rotation should be isolated. 

\end{appendix}

\end{document}